\documentclass[twoside,twocolumn,9pt]{article}
\usepackage{extsizes}
\usepackage[super,sort&compress,comma]{natbib} 
\usepackage[version=3]{mhchem}
\usepackage[left=1.5cm, right=1.5cm, top=1.785cm, bottom=2.0cm]{geometry}
\usepackage{balance}
\usepackage{mathptmx}
\usepackage{sectsty}
\usepackage{graphicx} 
\usepackage{lastpage}
\usepackage[format=plain,justification=justified,singlelinecheck=false,font={stretch=1.125,small,sf},labelfont=bf,labelsep=space]{caption}
\usepackage{float}
\usepackage{fancyhdr}
\usepackage{fnpos}
\usepackage{float}
\usepackage[english]{babel}
\addto{\captionsenglish}{%
  
}
\usepackage{array}
\usepackage{droidsans}
\usepackage{charter}
\usepackage[T1]{fontenc}
\usepackage[usenames,dvipsnames]{xcolor}
\usepackage{setspace}
\usepackage[compact]{titlesec}
\usepackage{hyperref}

\usepackage{epstopdf}

\definecolor{cream}{RGB}{222,217,201}

\begin{document}

\pagestyle{fancy}
\thispagestyle{plain}
\fancypagestyle{plain}{
\renewcommand{\headrulewidth}{0pt}
}

\makeFNbottom
\makeatletter
\renewcommand\LARGE{\@setfontsize\LARGE{15pt}{17}}
\renewcommand\Large{\@setfontsize\Large{12pt}{14}}
\renewcommand\large{\@setfontsize\large{10pt}{12}}
\renewcommand\footnotesize{\@setfontsize\footnotesize{7pt}{10}}
\makeatother

\renewcommand{\thefootnote}{\fnsymbol{footnote}}
\renewcommand\footnoterule{\vspace*{1pt}%
\color{cream}\hrule width 3.5in height 0.4pt \color{black}\vspace*{5pt}} 
\setcounter{secnumdepth}{5}

\makeatletter 
\renewcommand\@biblabel[1]{#1}
\renewcommand\@makefntext[1]%
{\noindent\makebox[0pt][r]{\@thefnmark\,}#1}
\makeatother 
\renewcommand{\figurename}{\small{Fig.}~}
\sectionfont{\sffamily\Large}
\subsectionfont{\normalsize}
\subsubsectionfont{\bf}
\setstretch{1.125} 
\setlength{\skip\footins}{0.8cm}
\setlength{\footnotesep}{0.25cm}
\setlength{\jot}{10pt}
\titlespacing*{\section}{0pt}{4pt}{4pt}
\titlespacing*{\subsection}{0pt}{15pt}{1pt}

\newcommand{\edd}[1]{\textcolor{red}{#1}}

\fancyfoot{}
\fancyfoot[LO,RE]{\vspace{-7.1pt}\includegraphics[height=9pt]{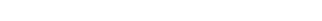}}
\fancyfoot[RO]{\footnotesize{\sffamily{1--\pageref{LastPage} ~\textbar  \hspace{2pt}\thepage}}}
\fancyfoot[LE]{\footnotesize{\sffamily{\thepage~\textbar\hspace{0.2cm} 1--\pageref{LastPage}}}}
\fancyhead{}
\renewcommand{\headrulewidth}{0pt} 
\renewcommand{\footrulewidth}{0pt}
\setlength{\arrayrulewidth}{1pt}
\setlength{\columnsep}{6.5mm}
\setlength\bibsep{1pt}

\makeatletter 
\newlength{\figrulesep} 
\setlength{\figrulesep}{0.5\textfloatsep} 

\newcommand{\topfigrule}{\vspace*{-1pt}%
\noindent{\color{cream}\rule[-\figrulesep]{\columnwidth}{1.5pt}} }

\newcommand{\botfigrule}{\vspace*{-2pt}%
\noindent{\color{cream}\rule[\figrulesep]{\columnwidth}{1.5pt}} }

\newcommand{\dblfigrule}{\vspace*{-1pt}%
\noindent{\color{cream}\rule[-\figrulesep]{\textwidth}{1.5pt}} }

\makeatother

\twocolumn[
  \begin{@twocolumnfalse} {
    \vspace{1em}
    \sffamily
    
    \noindent\LARGE{\textbf{Hydration of biologically relevant tetramethylammonium cation by neutron scattering and molecular dynamics$^\dag$}} \\
    \large{(9th November 2023)}
    
    \vspace{0.3cm}
    
    \noindent\large{Philip Mason$^{\ast}$\textit{$^{a}$}, Tomas Martinek\textit{$^{a}$}, Balázs Fábián\textit{$^{a \ddag}$}, Mario Vazdar\textit{$^{b}$}, Pavel Jungwirth\textit{$^{a}$}, Ondrej Tichacek\textit{$^{a}$}, Elise Duboué-Dijon\textit{$^{c}$}, and Hector Martinez-Seara$^{\ast}$\textit{$^{a}$}} \\
    
    \noindent\normalsize{Neutron scattering and molecular dynamics studies were performed on a concentrated aqueous tetramethylammonium (TMA) chloride solution to gain insight into the hydration shell structure of TMA, which is relevant for understanding its behavior in biological contexts of, e.g., properties of phospholipid membrane headgroups or interactions between DNA and histones. Specifically, neutron diffraction with isotopic substitution experiments were performed on TMA and water hydrogens to extract the specific correlation between hydrogens in TMA ($\mathrm{H_{TMA}}$) and hydrogens in water ($\mathrm{H_{W}}$). Classical molecular dynamics simulations were performed to help interpret the experimental neutron scattering data. Comparison of the hydration structure and simulated neutron signals obtained with various force field flavors (\textit{e.g.} overall charge, charge distribution, polarity of the CH bonds and geometry) allowed us to gain insight into how sensitive the TMA hydration structure is to such changes and how much the neutron signal can capture them. We show that certain aspects of the hydration, such as the correlation of the hydrogen on TMA to hydrogen on water, showed little dependence on the force field. In contrast, other correlations, such as the ion--ion interactions, showed more marked changes. Strikingly, the neutron scattering signal cannot discriminate between different hydration patterns. Finally, \textit{ab initio} molecular dynamics was used to examine the three-dimensional hydration structure and thus to benchmark force field simulations. Overall, while neutron scattering has been previously successfully used to improve force fields, in the particular case of TMA we show that it has only limited value to fully determine the hydration structure, with other techniques such as \textit{ab initio} MD being of a significant help.} \\

    }
\end{@twocolumnfalse} \vspace{0.6cm}
\bigskip
]

\renewcommand*\rmdefault{bch}\normalfont\upshape
\rmfamily
\section*{}
\vspace{-1cm}


\footnotetext{\textit{$^{a}$~Institute of Organic Chemistry and Biochemistry of the Czech Academy of Sciences, Flemingovo nám. 542, 160 00 Praha 6, Email: philip.mason@uochb.cas.cz, hseara@gmail.com}}
\footnotetext{\textit{$^{b}$~Department of Mathematics, Informatics, and Cybernetics, University of Chemistry and Technology Prague, Technická 5, 16628 Prague, Czech Republic}}
\footnotetext{\textit{$^{c}$~Laboratoire de Biochimie Théorique, CNRS, UPR 9080, Institut de Biologie Physico-Chimique, 13 rue Pierre et Marie Curie, 75005 Paris, France}}
\footnotetext{\textit{$^{\ddag}$~Present address: Department of Theoretical Biophysics, Max Planck Institute of Biophysics
, Max-von-Laue Straße 3, 60438 Frankfurt am Main, Germany}}

\footnotetext{\dag~Electronic Supplementary Information (ESI) available: ESI contains additional neutron scattering and AIMD analysis.}

\section{Introduction}

Tetramethylammonium (TMA) represents an important and ubiquitous motif in biological systems. It is found in the cellular membranes of almost all cells as the phosphorylcholine group in many phospholipids. Also, successive methylation of the amino acid lysine eventually results in a TMA functional group. This methylation is vital in histone--DNA binding and the epigenetic expression of DNA.\cite{cheung2005} Tetraalkylammonium salts are also powerful denaturants\cite{Hippel1965} and are widely used as phase transfer catalysts\cite{Brandstrom1977} and are also seen frequently in ionic liquids.\cite{Mustain2023}

TMA is one of the simplest and most spherical representatives of the so-called ``hydrophobic ions'', as compared to more complex variants, such as tetraphenylphosphonium or tetrabutylammonium. TMA has short hydrophobic chains, so an intriguing question arises. How much does its hydration properties differ from those of a hypothetical perfectly spherical large ion?\cite{Turner1990,Polydorou1997,Lang1990} Small spherical alkali cations have a fairly generic hydration structure, with the first hydration shell water oxygen atom pointing towards the cation and the hydrogen atoms away. This orientation is the strongest for lithium, getting weaker upon moving down the periodic table. How does the solvation shell structure change by the time we get to a large cation like TMA? Also, given the hydrophobic character of the methyl groups, does hydrophobicity-driven cation--cation aggregation take place in aqueous TMA solutions?\cite{Bhowmik2014}

Neutron scattering experiments have provided insight into TMA hydration structure in TMA chloride or bromide solutions.\cite{Turner1990,Turner1992,Turner1995,Polydorou1997,Nilsson2016} In aqueous solutions, the neutron scattering signal is dominated by water--water correlations. The structural correlations of interest, such as those between TMA and water nuclei, are thus hidden among the dominant water--water correlations. A way to deal with this problem is to employ neutron diffraction with isotopic substitution (NDIS).\cite{Jong1996,Mason2006_10.1021/ja0613207,Mason2006_10.1088/0953-8984/18/37/004,Neilson2001} This method takes advantage of the fact that different isotopes have different neutron scattering properties and relies on the assumption that the mass of a nucleus does not affect the structure of the solution. This assumption of neglecting nuclear quantum effects has proven fairly robust even for materials with the largest isotope effects. Even H$_2$O and D$_2$O vary in number density by less than half a percent despite substituting two-thirds of the nuclei in the system. Still, hydrogen and deuterium are excellent nuclei to use in NDIS. Both are easily experimentally available and have among the largest neutron scattering contrasts for any element. The technique requires two identical solutions to be prepared, which differ only in isotopic concentration of one nucleus. As the structure of these solutions is assumed to be identical, the subtraction of their diffraction patterns cancels out for all components unrelated to the substituted nucleus. With its 12 identical hydrogen nuclei and the large contrast between H and D isotopes, the TMA ion is an ideal candidate to be studied by neutron scattering.\cite{Turner1990, Turner1992, Turner1995}
However, even if NDIS makes isolating a few structural correlations of interest possible, not all pairwise distributions are equally easy to understand. The intuitive way of examining TMA hydration is to look at the spatial distribution of water oxygen atoms around the central TMA nitrogen atom. Unfortunately, NDIS cannot provide this specific structure factor due to lack of suitable oxygen isotopes. The relatively easily experimentally accessible $\mathrm{H_{TMA}}$-$\mathrm{H_{W}}$ structure factor corresponds neither to the center of the TMA nor that of the water molecule, which makes intuitive interpretations of the data significantly more complicated. 

Despite these limitations, early NDIS experiments \cite{Turner1990, Turner1992, Turner1994, Turner1995} were interpreted as evidence of apolar hydration around TMA, with an edge-on orientation of water molecules. 
More recently, Monte--Carlo-based empirical potential structure refinement (ESPR) simulations were used to assist the interpretation of the neutron scattering signal, confirming the apolar character of TMA hydration but suggesting that water arranges tetrahedrally around TMA.\cite{Nilsson2016} In the past years, molecular dynamics simulations have proven to be highly valuable to help further the interpretation of neutron scattering signals,\cite{Pluharova2014,Martinek2018} which is also the strategy adopted in the present work.
In particular, we performed NDIS experiments on a 2~m TMACl solution, allowing us to extract a single structural correlation between hydrogens of the cation $\mathrm{H_{TMA}}$ and those of water $\mathrm{H_{W}}$. We employed DFT-based ab initio molecular dynamics (AIMD) and force field molecular dynamics (FFMD) simulations to assist in interpreting the experimental signal. Simulations with various force fields were performed to test the sensitivity of the measured signal to various changes in the hydration structure, which turned out to be the key to understanding the exact amount of information contained in the neutron scattering signal and avoiding over-interpretations. In addition, we obtained new insights into optimizing force fields to better represent the hydration shells around solutes.

\section{Methods}

\subsection{Neutron scattering measurements}

\begin{figure}[ht!]
\centering
  \includegraphics[width=9cm]{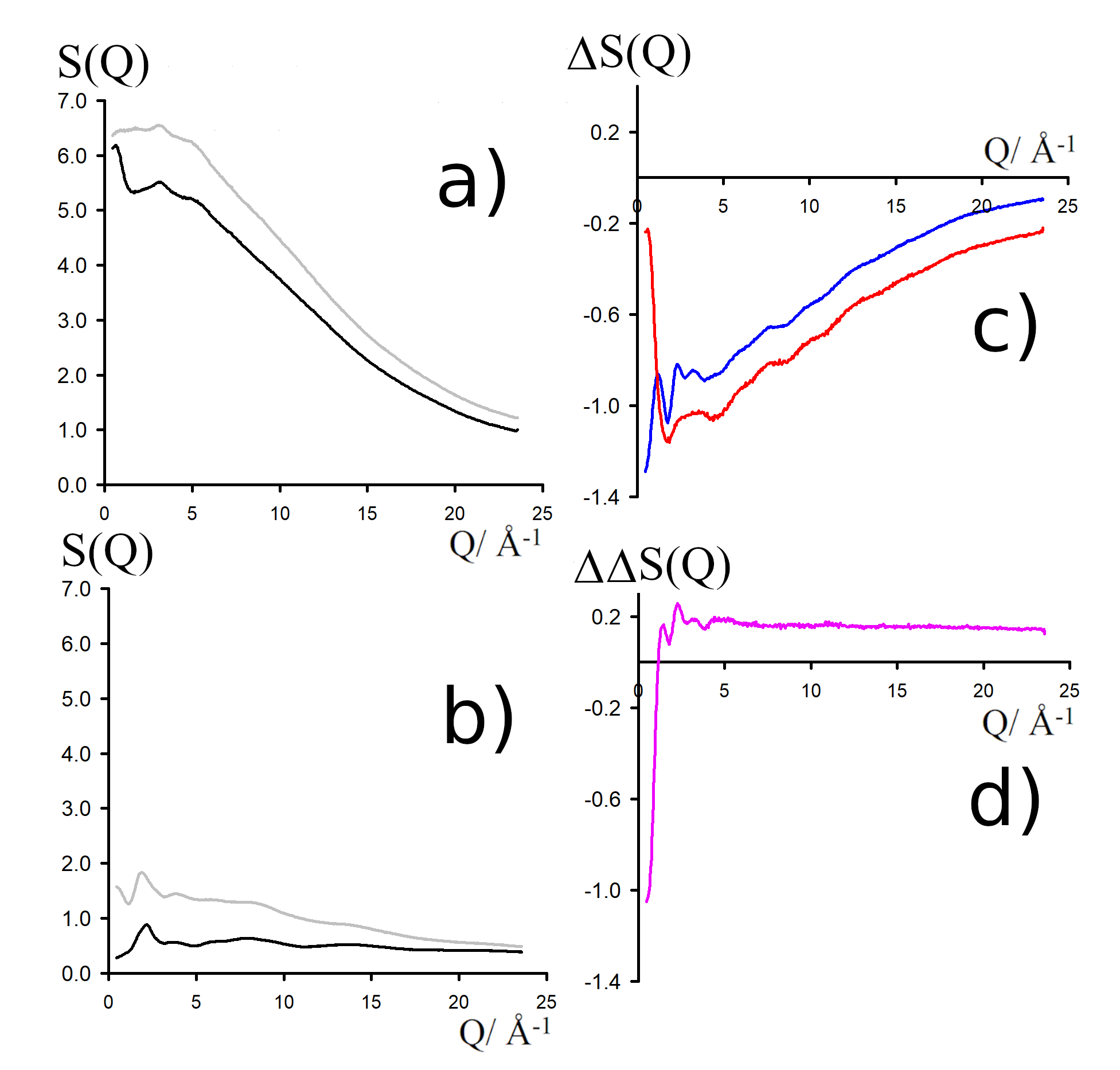}
  \caption{a) Total diffraction patterns for H$_2$O solutions of d-TMACl(black) and h-TMACl(grey). b) Total diffraction patterns for D$_2$O solutions of d-TMACl(black) and h-TMACl(grey). c) First order differences $\Delta S^\mathrm{X_{H_2O}}_\mathrm{H_{non}}(Q)$ (red line, obtained by the difference of the two diffraction patterns shown in a) and $\Delta S^\mathrm{X_{D_2O}}_\mathrm{H_{non}}(Q)$ (blue line, obtained by the difference of the two diffraction patterns shown in b). d) Second order difference $140.8 \cdot \left ( S_\mathrm{H_{TMA}H_W}(Q) -1 \right)$, obtained through the difference of the two first-order differences shown in c).}
  \label{fgr:TotalDifraction}
\end{figure}

NDIS measurements were performed using the D4C diffractometer at the nuclear reactor at the Institut Laue--Langevin in Grenoble, France.\cite{Fischer2002} Four chemically identical solutions of 2~m TMACl in water were prepared, which differed only in H/D substitution on the TMA and H/D substitution on water. The four diffraction patterns (Fig.~\ref{fgr:TotalDifraction}a,b) were recorded for about 2~hrs for each D$_2$O solution and for 4~hrs for each H$_2$O solution. The results were then corrected for multiple scattering and absorption and normalized against a standard vanadium scatterer.\cite{Herdman1992}

Taking the difference between the diffraction patterns associated with solutions that differ only by the H/D substitution on TMA (both in D$_2$O and H$_2$O solutions) yields the first-order differences $\Delta S^\mathrm{X_{D_{2}O}}_\mathrm{H_{non}}(Q)$ and $\Delta S^\mathrm{X_{H_{2}O}}_\mathrm{H_{non}}(Q)$ (Fig.~\ref{fgr:TotalDifraction}c), that report on the correlation between non-exchangeable H on TMA and every other atom (X) in the system. They are respectively defined as (in units of mbarns):

\begin{equation}
  \label{eqn:Q1stH2O}
  \begin{split}
  \Delta S^\mathrm{X_{D_{2}O}}_\mathrm{H_{non}}(Q) 
            & = 90.2 \cdot S_\mathrm{H_{TMA}H_{W}}(Q) + 39.3 \cdot S_\mathrm{H_{TMA}O}(Q) \\
            & + 6.5  \cdot S_\mathrm{H_{TMA}C}(Q)     + 2.3  \cdot S_\mathrm{H_{TMA}N}(Q) \\
            & + 2.3  \cdot S_\mathrm{H_{TMA}Cl}(Q)    + 4.3  \cdot S_\mathrm{H_{TMA}H_{TMA}}(Q) - 144.8,
  \end{split}
\end{equation}

\begin{equation}
  \label{eqn:Q1stD2O}
  \begin{split}
  \Delta S^\mathrm{X_{H_{2}O}}_\mathrm{H_{non}}(Q)
            & = -50.6 \cdot S_\mathrm{H_{TMA}H_{W}}(Q) + 39.3 \cdot S_\mathrm{H_{TMA}O}(Q) \\
            & + 6.5   \cdot S_\mathrm{H_{TMA}C}(Q)     + 2.3  \cdot S_\mathrm{H_{TMA}N}(Q) \\
            & + 2.3   \cdot S_\mathrm{H_{TMA}Cl}(Q)    + 4.3  \cdot S_\mathrm{H_{TMA}H_{TMA}}(Q) - 4.0.
  \end{split}
\end{equation}
These prefactors were calculated from the atomic concentration and neutron scattering lengths of the various elements in the system by standard literature methods.\cite{Enderby1975}

The difference between eq.~\ref{eqn:Q1stH2O} and eq.~\ref{eqn:Q1stD2O} yields the second order difference $\Delta\Delta S_\mathrm{H_{non}}(Q)$ (Fig.~\ref{fgr:TotalDifraction}d), which reports on the single correlation between the TMA non-exchangeable H atoms and the water H atoms.

\begin{equation}
  \label{eqn:Q2nd}
  \Delta\Delta S_\mathrm{H_{non}}(Q) = \Delta S_\mathrm{H_{non}}^\mathrm{X_{D_{2}O}}(Q) - \Delta S_\mathrm{H_{non}}^\mathrm{X_{H_{2}O}}(Q) = 140.8 \cdot \left ( S_\mathrm{H_{TMA}H_W}(Q) -1 \right)
\end{equation}

This function provides a useful internal consistency check for the accuracy of the solutions and the multiple scattering and absorption corrections performed on the data. Due to the large inelastic scattering of $^1$H and the Placzek effect,\cite{Mason2006NeutronDiffractionSimulation} hydrogen-containing samples always present a dominant background. The higher the atomic concentration of $^1$H, the larger the effect, primarily visible in light water samples such as in Fig.~\ref{fgr:TotalDifraction}a. The effect is greatly diminished for the heavy water samples, as shown in Fig.~\ref{fgr:TotalDifraction}b.
The amount of inelastic scattering is largely determined by the number of $^1$H nuclei per unit volume, so the first-order differences (Fig.~\ref{fgr:TotalDifraction}c) should have exactly the same Placzek background, which should vanish completely in the second order difference (Fig.~\ref{fgr:TotalDifraction}d). The largely constant value of the second-order difference in Fig.~\ref{fgr:TotalDifraction}d indicates the absence of a visible background and proves that the four solutions were prepared with the same chemical composition.


\subsection{Force field molecular dynamics (FFMD) simulations}

We performed classical FFMD simulations of seven systems comprising 50 TMA cations or, for comparison, 50 neutral neopentane molecules matching the experimental concentration of 2~m. Force field details for each system are available in Table~\ref{tbl:TMAmodels}. Systems containing TMA were neutralized with chloride counterions of the CHARMMCl\_o type \cite{Chandrasekhar1984} ($-1$~charge) or CL\_2s type 13 ($-0.75$~scaled charge). Each system contained 1388 TIP3P water molecules with Lennard-Jones parameters also on the hydrogens \cite{MacKerell1998}. All systems were assembled using GROMACS 2021.2 and 2021.5 tools.\cite{Abraham2015}

For the TMA moiety, several parametrizations used in the head group of lipids are available in the literature, which differ mainly in their partial charges on the three atom types of the TMA group. First, we employed the CHARMM36 parameters with the full +1 charge (denoted \texttt{CHARMM}). Next, we used a scaled variant, \texttt{prosECCo}, where charges were scaled by 0.75. This physically based Electronic Continuum Correction (ECC) approach accounts for electronic polarization in a mean-field way by scaling charges\cite{Leontyev2011}. ECC has been shown to improve significantly the description of ion pairing behavior in solution. \cite{DuboueDijon2020,kostal2023} Based on this scaled charge model with an overall charge of $+0.75$, we designed three additional variants with different charge distributions but the same overall $+0.75$ charge: 1) \texttt{low CH dipole}, where partial charges are reduced; 2) \texttt{central-N}, where the whole charge is concentrated on the nitrogen; 3) \texttt{surface-H}, where all charge is redistributed on the hydrogens (see Table~\ref{tbl:TMAmodels}). All these TMA models have the same Lennard-Jones parameters, as prescribed in CHARMM36. 

For comparison, we also performed simulations with the neutral neopentane with the same geometry as TMA using the CHARMM36 force field. Note that this compound is highly hydrophobic and aggregates quickly when mixed in water. Finally, we modeled a single charged sphere similar to the coarse-grained model of TMA, \texttt{center-bead}. In this coarse-grained-like simulations, water still has an atomistic resolution while the entire tetramethylammonium group is replaced by a single ``extended atom''. The size of this so-called bead is set to match the one of TMA in the first peak of its radial distribution with the surrounding water.\cite{Bennun2009}

\begin{table*}
\small
  \caption{\ TMA and neopentane simulation models used in this study. In parenthesis, the used CHARMM36 atoms types\cite{Klauda2010UpdateCHARMMAll} are listed.}
  \label{tbl:TMAmodels}
  \begin{tabular*}{\textwidth}{@{\extracolsep{\fill}}|c|c|ccc|c|}
    \hline
    \textbf{Model}   & \textbf{Geom.} & \multicolumn{3}{c|}{\textbf{Atom (atom type) partial charge}} &  \textbf{Overall charge}    \\
   
    &  & N (NTL) & C (CTL5) & H (HL) &  \\  \hline
    \texttt{CHARMM} & TMA & --0.60 & --0.35 & 0.25 & +1.00 \\
    \texttt{prosECCo} & TMA & --0.61 & --0.35 & 0.23 & +0.75 \\
    \texttt{low CH dipole} & TMA & --0.05 & --0.10 & 0.10 & +0.75 \\
    \texttt{center-N} & TMA & +0.75 & 0.00 & 0.00 & +0.75 \\
    \texttt{surface-H} & TMA & 0.00 & 0.00 & 0.0625 & +0.75 \\
    \hline
    \multicolumn{6}{|c|}{}   \\
    \hline
    \multicolumn{2}{|c|}{} &\multicolumn{3}{c|}{ \texttt{center-bead} ($\sigma = 0.550$~nm, $\epsilon = 0.83680$~kJ/mol)} &   \\ \hline
     \texttt{center-bead} & sphere & \multicolumn{3}{c|}{ +0.75}  & +0.75 \\
     \hline
    \multicolumn{6}{|c|}{}   \\
    \hline
    \multicolumn{2}{|c|}{} & C (CT) & C (CT3) & H (HA3) &  \\  \hline
    \texttt{neopentane} & TMA & 0.00 & --0.30 & 0.10 & 0.0 \\ \hline
  \end{tabular*}
\end{table*}

All systems were simulated with GROMACS 2021, \cite{Abraham2015} using simulation parameters as provided by CHARMM-GUI.\cite{Lee2016} The LINCS and Settle algorithms were used to constrain the geometry of TMA hydrogens \cite{Hess1997} and water molecules, respectively, allowing the use of 2~fs time step in the simulations. \cite{Miyamoto1992} The coordinates were saved every 10~ps for each 2~$\mu$s simulation. For neopentane that aggregates due to its hydrophobic nature, we saved coordinates every 20~fs to gather enough water orientation statistics to study its coordination before aggregation dominates (\textit{i.e.}, $\approx$ 0.5~ns). Simulations were performed in the isothermal--isobaric ($NpT$) ensemble. Temperature was controlled using a Nos\'e--Hoover thermostat \cite{Nose1984} with a time constant of 1~ps and a reference value of 310~K, and a constant pressure of 1~bar was maintained by an isotropically coupled Parrinello--Rahman \cite{Parrinello1981} barostat with a time constant of 5~ps. Van der Waals interactions were treated using a cutoff of 1.2~nm with a force-switch at 1.0~nm using Verlet cutoff scheme \cite{Pall2013} for neighbors. Long-range Coulomb interactions were accounted for using particle mesh Ewald (PME) with a 1.2~nm cutoff as implemented in GROMACS.\cite{Abraham2015} 

\subsection{\textit{Ab Initio} molecular dynamics (AIMD) simulations}

As a benchmark for the TMA solvation structure, we complemented the neutron diffraction experiments and the force field-based simulations with Born--Oppenheimer \textit{ab initio} molecular dynamics simulations (AIMD) of a single TMA cation with 64 water molecules under periodic boundary conditions. The large computational cost of AIMD simulations precludes using larger systems, such as those in FFMD. The present system was not neutralized by any counterion. Force field molecular dynamics was used to preequilibrate the system and to prepare the initial configurations for the subsequent AIMD simulation, as follows, with a constant pressure simulation used to estimate the average size of the cubic simulation cell of 12.552~{\AA}. A constant volume simulation was then used to prepare 10 initial configurations separated by 2~ns and equilibrated for 1~ns using FFMD. From these 10 equilibrated structures, AIMD simulations were performed using the generalized gradient approximation revPBE DFT functional\cite{Perdew1996,Zhang1998CommentGeneralizedGradient,Perdew1998PerdewBurkeErnzerhof} with the D3 dispersion correction\cite{Grimme2006,Grimme2010,Grimme2011}. Core electrons were replaced by GTH pseudopotentials,\cite{Goedecker1996,Krack2005}, and the triple-$\zeta$ basis set TZV2P with polarization functions was used for valence electrons.\cite{VandeVondele2007} A cutoff of 400~Ry was used for the auxiliary plane wave basis set in the GPW method.\cite{Lippert1997} The system was first equilibrated using AIMD Langevin dynamics for at least 16~ps with a damping constant of $\gamma$ = 0.02~ps$^{-1}$\cite{gunsteren1982}. During the production simulation, the temperature was set to 300~K via a global CSVR thermostat with a time constant of 1~ps.\cite{Bussi2007} To enhance sampling, ten 50~ps parallel AIMD simulations totaling 500~ps of trajectory were used for further structural analysis. All AIMD simulations were performed at constant volume, using the CP2K program package, version 7.1.\cite{Hutter2014,kuhne2020,Vandevondele2005}

\subsection{Density maps}

Data from FFMD and AIMD simulations were processed with the help of an in-house developed software for unbiased alignment and density analysis. The analysis included atoms found within a 10~\AA{} of the nitrogen atom in TMA, covering all TMA molecules in the system and all simulation frames. The TMA ``neighborhoods'' were aligned in two steps. First, a representative conformation was found for each simulation, and then the neighborhoods were aligned to that conformation using the positions of carbon atoms. In the case of \texttt{center-bead} approximation of TMA, the neighboring oxygen/chlorine atoms were used for the alignment instead. The alignment was performed similarly as in Ref. \citenum{Nguyen2021Resolvingequalnumber}, that is, using a permutation-based unconstrained alignment, such that the reference atoms used for the alignment (TMA carbons or oxygens and chlorines within its first solvation shell) do not need to be labeled and sorted. This way, the thermal noise is uniformly distributed in the density maps of all systems.

\section{Results and Discussion}

\begin{figure}[ht!]
\centering
  \includegraphics[width=8cm]{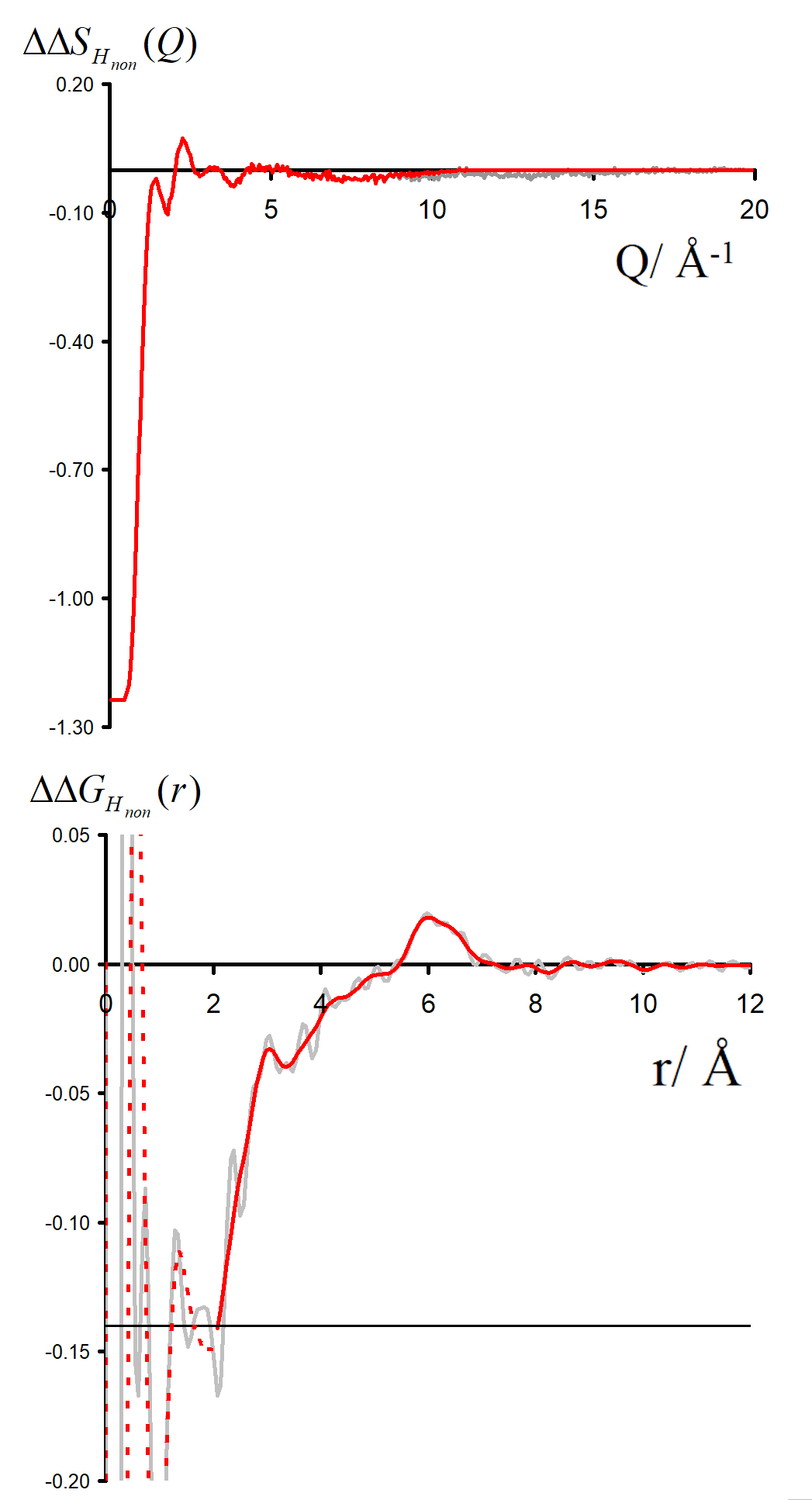}
  \caption{Upper the function $\Delta\Delta S_\mathrm{H_{non}}(Q)$ (grey) and lower the direct Fourier transform of this function $\Delta\Delta G_\mathrm{H_{non}}(r)$ (grey). Shown in red is the function $\Delta\Delta S_\mathrm{H_{non}}(Q)$ used for the rest of this paper, terminating the data using a windows function up to $\simeq 10$~{\AA}$^{-1}$. The lower red function is the real space version of the upper red function.}
  \label{fgr:Qr2ndDifference}
\end{figure}

Neutron scattering patterns were obtained for four identical TMACl solutions that differed only by their isotopic composition on the nonexchangeable H of TMA and the exchangeable water H (see Methods section). While the diffraction patterns of the solution are dominated by the signal coming from the water signal, taking differences between pairs of solutions leads to the cancellation of the signal part that does not depend on the isotopic composition. Hence, first-order differences (see Methods) report the TMA(H/D) correlation to every other atom in the system. While rich in information, this signal still contains all the intramolecular correlation peaks, which hides the information about hydration (see ESI). Hence, we proceeded to obtain the double difference signal $\Delta\Delta S_\mathrm{H_{non}}(Q)$ (see Methods and Fig.~\ref{fgr:Qr2ndDifference}), which reports on the single correlation between $\mathrm{H_{TMA}}$ and $\mathrm{H_{W}}$. It thus directly probes the TMA hydration structure and is much easier to interpret than the total diffraction patterns. The obtained $\Delta\Delta S_\mathrm{H_{non}}(Q)$ exhibits neatly resolved features below 10~\AA\, which characterize TMA hydration.
Interpretation of the signal tends to be more intuitive in direct space. Hence, we computed the Fourier-transform $\Delta\Delta G_\mathrm{H_{non}}(r) = \mathcal{F} \left [\Delta\Delta S_\mathrm{H_{non}}(Q) \right] (t)$, which is directly related to the single pair-correlation function $g_\mathrm{H_{TMA}H_W}$:

\begin{equation}
  \label{eqn:G2nd}
  \Delta\Delta G_\mathrm{H_{non}}(r) = \Delta G_\mathrm{H_{non}}^\mathrm{X_{D_{2}O}}(Q) - \Delta G_\mathrm{H_{non}}^\mathrm{X_{H_{2}O}}(Q)
                                     = 140.8 \cdot (g_\mathrm{H_{TMA}H_W}(r) - 1)
\end{equation}

$\Delta\Delta G_\mathrm{H_{non}}(r)$ shown in Fig.~\ref{fgr:Qr2ndDifference} presents a characteristic shoulder around 3--4~{\AA}, and a peak at 6~\AA. While direct molecular interpretation of these features is not straightforward, the same pair correlation function can be easily computed from molecular dynamics simulations with two objectives in mind. First, we want to assist in interpreting the neutron diffraction signal and obtain a molecular-level picture of the TMA hydration. Second, we want to investigate how sensitive TMA hydration is to variations in the intermolecular interactions, and whether the resulting patterns in hydration structure would be distinguishable in the neutron signal. To this aim, we performed FFMD simulations of a TMACl solution, at the same 2~m concentration as used in the neutron scattering experiment, using different force field variants (see Methods, Table~\ref{tbl:TMAmodels}), characterizing in each case the hydration structure and the associated neutron signal. 

\subsection{What is the effect of overall charge on TMA solvation?}

TMA is a large cation with a relatively low charge density. Hence, a question arises on how different its hydration structure is from that of neutral solutes, such as neopentane (beyond the absence of counterions and possible aggregation, as discussed below). In addition, within standard force fields, TMA, as intuitively expected, is assigned a global charge of +1. However, electronic polarization, which further screens interactions between ions, is missing in FFMD simulations using non-polarizable force fields, which may lead to artefacts such as excessive ion pairing.\cite{Leontyev2011,DuboueDijon2020} A mean-field strategy to implicitly account for electronic polarization in FFMD simulations is the Electronic Continuum Correction (ECC) approach, which is mathematically equivalent to scaling partial charges by a factor $1/\sqrt{\epsilon _\infty} \simeq 0.75$, where $\epsilon _\infty$ is the high-frequency dielectric constant of water. Hence, we designed an ECC version of the TMA force field (denoted as \texttt{prosECCo}, see Table~\ref{tbl:TMAmodels}), with an overall charge of +0.75.\cite{DuboueDijon2020,Nencini2022Proseccopolarizationreintroduced} Here, we compared the structure of the solution simulated using this scaled-charge force field with that obtained using the standard full charge CHARMM36 force field, as well as to that of a solution containing electrically neutral neopentane molecules for comparison. Note that neopentane is insoluble and starts precipitating fast in the simulations. Consequently, we used only the first 30~ps (using 1500 frames, 20~fs between frames) of this simulation for analysis.

\begin{figure*}[ht!]
\centering
  \includegraphics[width=18cm]{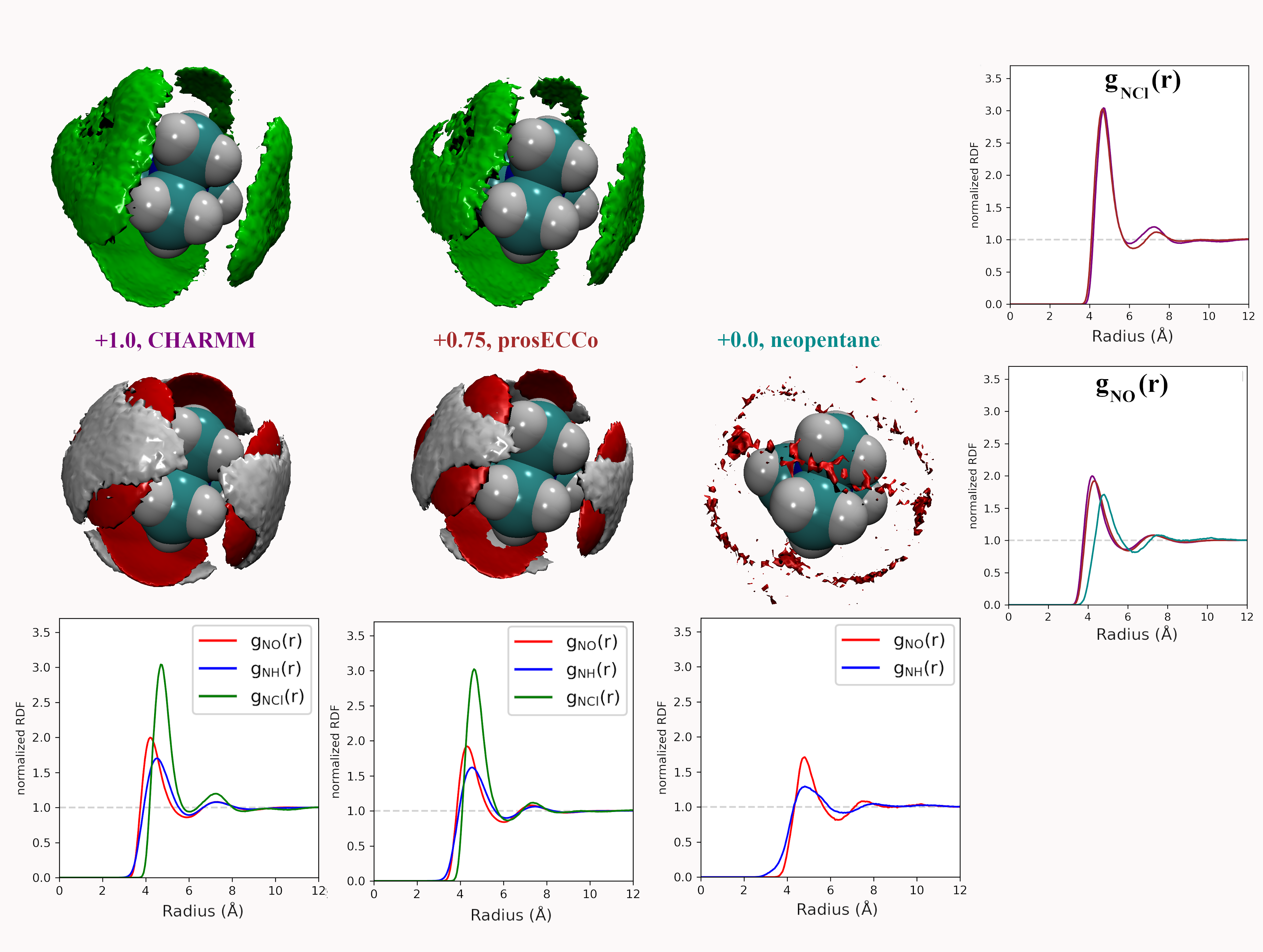}
  \caption{Comparison charge effects in first solvation shell of TMA geometries. The first column shows charge +1.0 TMA (\texttt{CHARMM}), the second column +0.75 TMA (\texttt{prosECCo}), and the third column (\texttt{neopentane}). The first row shows chloride ions and the lower oxygen and hydrogen atoms. In all density maps, the density of chloride ion (green) is six times bulk density, oxygen (red) is three times bulk density, and hydrogen (white) is two times bulk density. The lower row shows the RDFs from the central atom for each model with oxygen (red), hydrogen (blue), and chloride (green). The fourth column compares the N-Cl$^-$ and N/C-O RDFs for the three models.}
  \label{fgr:MapsCharge}
\end{figure*}

For each simulation, we obtained radial distribution functions from the central nitrogen atom to the surrounding O, $\mathrm{H_{W}}$, and Cl$^-$, as well as density maps of chloride, water oxygen, and water hydrogen around TMA (see Fig.~\ref{fgr:MapsCharge}). Fig.~\ref{fgr:MapsCharge} clearly shows that changing the charge from +1 to +0.75 has only a minor effect on the hydration structure, resulting in nearly identical radial distribution functions. If we view the TMA ion as a tetrahedron with four faces, six edges, and four corners, the water oxygen atoms (as well as the chloride counterion) are located more at the faces, less at the edges, and not at all at the corners. The amount of TMA$^+$--Cl$^-$ ion pairing is also very similar for the two systems, exhibiting only a small excess of solvent-shared ion pairs in the full charge force field compared to that with scaled charges (see second peak at 6--8~\AA\ in $g_\mathrm{NCl}(r)$). This is contrary to what was previously observed for small monovalent cations such as Li$^+$ \cite{Pluharova2013}, and with divalent ions such as Ca$^{2+}$ \cite{Kohagen2014}, where charge scaling changed qualitatively the number of ion pairs. In contrast, due to the low charge density of TMA, only minor polarization effects are observed for ion hydration and pairing.

If we now compare TMA hydration with that of neutral neopentane, the main difference is due to a size effect with the hydration layer shifted slightly further away from the central atom. The orientation of the water molecules is also different -- in neopentane, the water OH bonds point very slightly towards the central atom; in contrast, in TMA, the water dipole orients towards the ion, and the OH points slightly outwards. Hence, while much less strongly oriented than around smaller ions such as lithium, the hydration of TMA still appears in these simulations significantly different from that of a hydrophobic solute. Note that the observations discussed are robust, despite neopentane statistics being worse due to less sampling due to their strong clustering propensity.

\subsection{Effect of charge distribution on TMA solvation}

\begin{figure*}[ht!]
\centering
  \includegraphics[width=18cm]{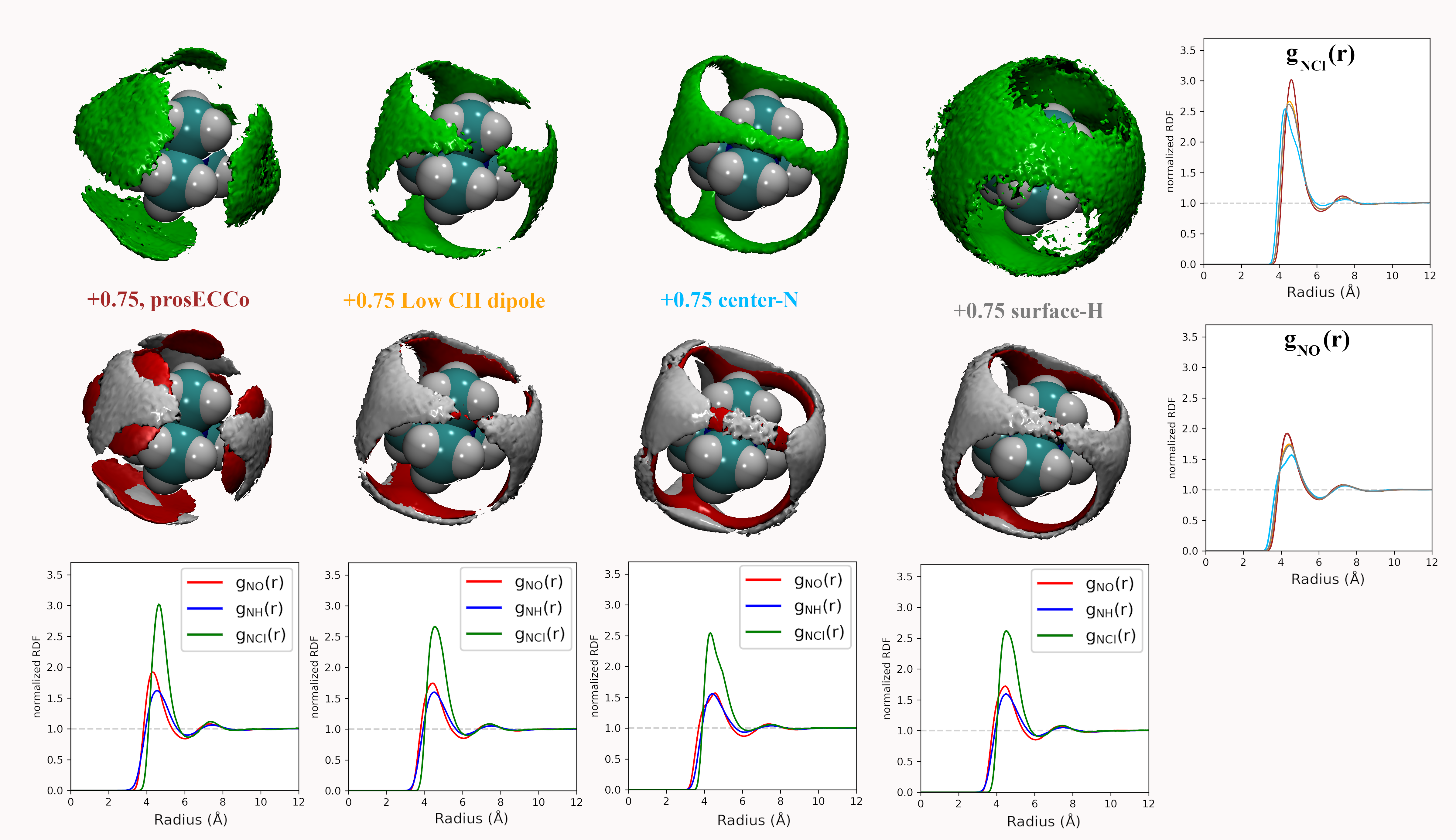}
  \caption{For all density maps, the density of chloride ion (green) is 6x bulk density, oxygen (red) is 3x bulk density and hydrogen (white) is 2x bulk density. All the TMA cations in this figure have a +0.75 charge. Left column, \texttt{prosECCo} force field, center left, the polarity of the CH bond is comparable to that of neopentane, with the remaining charge on the central nitrogen. Center right, all of the charge of the ion is on the central nitrogen, and far right, all the charge is spread evenly on the hydrogen atoms. Lower are shown the RDFs from the central atom of the TMA to oxygen (red), hydrogen (blue), and chloride (green). Right are shown the same RDFs but grouped for each four FF. Upper for the chloride ion and lower for the oxygen atom.
  \label{fgr:MapsDistribution}
  }
\end{figure*}

For a given overall charge, we investigated how the charge distribution within the TMA impacted its hydration structure and its propensity to form TMA--Cl ion pairs. For water as strong hydrogen bonding moiety, the partial charge on the central oxygen is about $-$0.8 with +0.4 on the hydrogen, while for TMA, the partial charges in most force fields are about $-$0.3 on the carbon and +0.2 on the hydrogen.
Starting from the scaled charge \texttt{prosECCo} model with an overall TMA charge of +0.75, we compared four different charge distributions -- the \texttt{prosECCo} force field (with results presented above), a \texttt{low CH dipole} force field variant where the charges on the methyl C and H atoms are respectively $-$0.1 and +0.1, a \texttt{center-N} force field where all the charge is placed on the central nitrogen, and a \texttt{surface-H} force field where the charge is equally distributed over the surface hydrogens of the TMA only. Note that while the charge distribution is rather different between the \texttt{center-N} and \texttt{surface-H} models, they both result in a very low polarity of the C--H bonds and zero or very small charge on hydrogens. In each of the four cases, we characterized the hydration structure and ion pairing via the radial distribution functions $g_\mathrm{NH}(r)$, $g_\mathrm{NO}(r)$ and $g_\mathrm{NCl}(r)$ and the density maps of $\mathrm{H_{W}}$, O and Cl around TMA (Fig.~\ref{fgr:MapsDistribution}). 

Our calculations show that changes in the charge distribution within TMA (see Fig.~\ref{fgr:MapsDistribution} have a much larger effect on hydration structure than the reduction of the overall charge of the ion from +1.0 to +0.75 (see Fig.~\ref{fgr:MapsCharge}. Namely, lowering the charge on the surface hydrogens qualitatively changes the hydration pattern. The density maps (Fig.~\ref{fgr:MapsDistribution}) clearly show that the water oxygens (as well as the chloride counterions, which follow the same trends) then move to the center of the faces of the TMA tetrahedron (i.e., away from the H atoms), forming bridges over the tetrahedron edges. This is manifested in the radial distribution functions $g_\mathrm{NH}(r)$, $g_\mathrm{NO}(r)$ and $g_\mathrm{NCl}(r)$ as a subtle increase in the bimodality of the first peak for $g_\mathrm{NO}(r)$ and $g_\mathrm{NCl}(r)$ in \texttt{center-N}. At the same time, changing the charge distribution does not significantly change the orientation of the hydration water OH bonds (see Fig.~\ref{fgr:MapsDistribution}).

\subsection{Effect of molecular geometry on TMA solvation}

\begin{figure*}[ht!]
\centering
  \includegraphics[width=18cm]{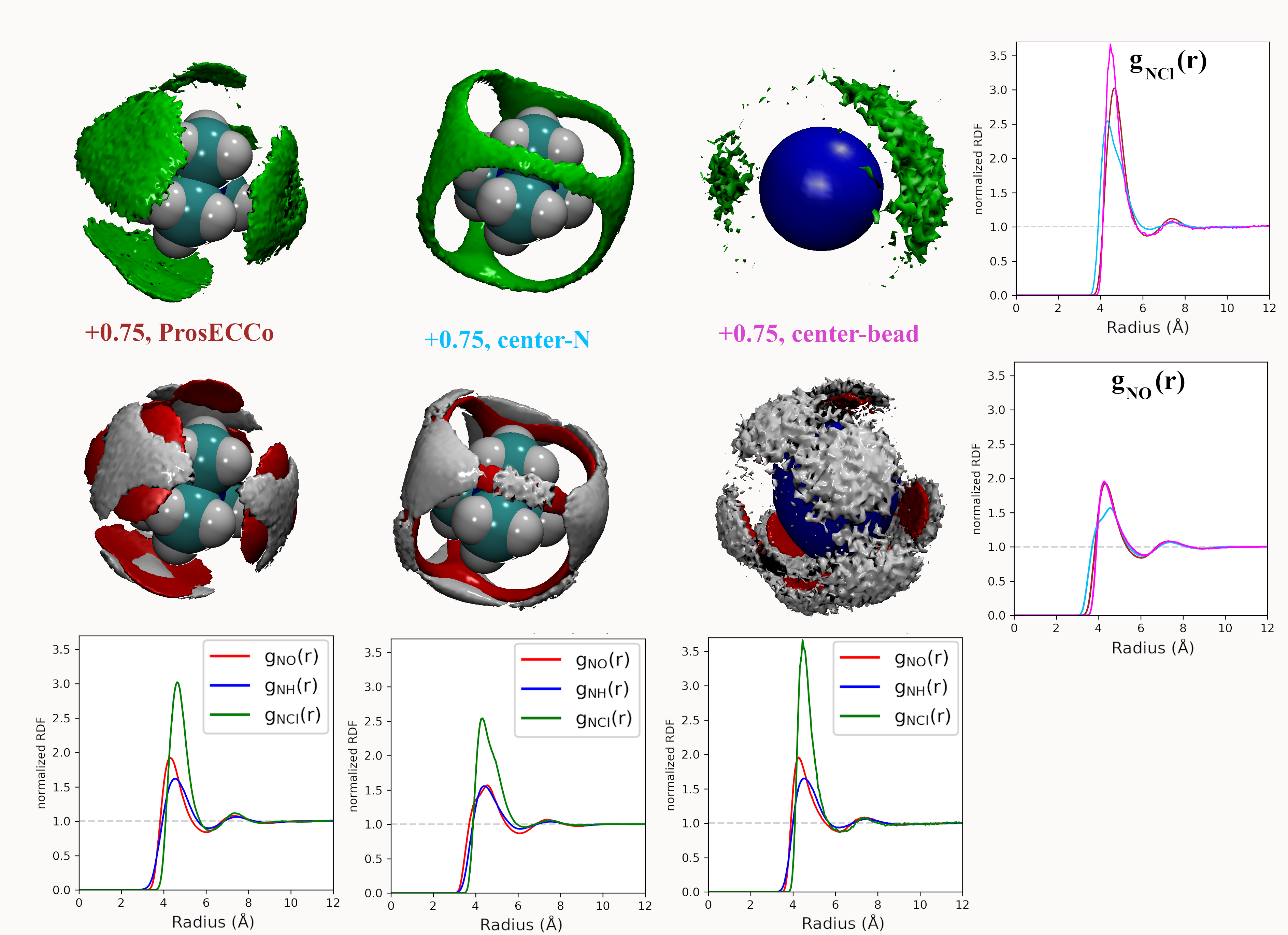}
  \caption{For all density maps, density of chloride ion (green) is 6x bulk density, oxygen (red) is 3x bulk density, and hydrogen (white) is 2x bulk density. All the ions in this figure have +0.75 charge. Left column, \texttt{prosECCo} force field, center all of the charge of the ion is on the central nitrogen, and right, as center but all the tetrahedral structure of the ion has been replaced by a single VDW sphere. Lower are shown the RDFs from the central atom of the TMA to oxygen (red), hydrogen (blue), and chloride (green). Right are shown the same RDFs but group for each three FFs. Upper for the chloride ion and lower for the oxygen atom.}
  \label{fgr:MapsGeometry}
\end{figure*}

Coarse-grained models of TMA reduce the molecular geometry to a single spherical bead. Is this loss of molecular structure important or, in other words, how much does TMA behave as a simple charged sphere? To examine this issue we performed a set of three simulations with the all-atom \texttt{prosECCo} force field, the \texttt{center-N} force field variant where the charge is localized on the central N atom, and the coarse-grained force field where all the charge is at the center of a single spherical bead. Again, the density maps for O, $\mathrm{H_{W}}$ and Cl were calculated around TMA, as well as the radial distribution functions $g_\mathrm{NH}(r)$, $g_\mathrm{NO}(r)$ and $g_\mathrm{NCl}(r)$, see Fig.~\ref{fgr:MapsGeometry}.

Interestingly, the all-atom \texttt{prosECCo} model is more similar in terms of the NO and N-$\mathrm{H_{W}}$ radial distribution functions to the coarse-grained charged sphere than to the \texttt{center-N} model (Fig.~\ref{fgr:MapsGeometry}). Notably, the radial distribution functions of the \texttt{center-N} model are much more bimodal than the two others due to the combination of structural arrangements around the TMA tetrahedral features of faces, edges, and corners. This means that much of the TMA hydration structure reflects the constraints imposed by the charged sphere of a given size on the water H-bond network. At the same time, the atomistic TMA model is strikingly different from a simple charged sphere when looking at its interaction with the chloride counterion. In the former, the counterions adopt a tetrahedral geometry at the center of the faces similar to the oxygen atoms, while the latter tends to form linear Cl-TMA-Cl structures (Fig.~\ref{fgr:MapsGeometry}). 

\subsection{Comparison of neutron scattering data to MD simulations}

\begin{figure}[ht!]
\centering
  \includegraphics[width=7.5cm]{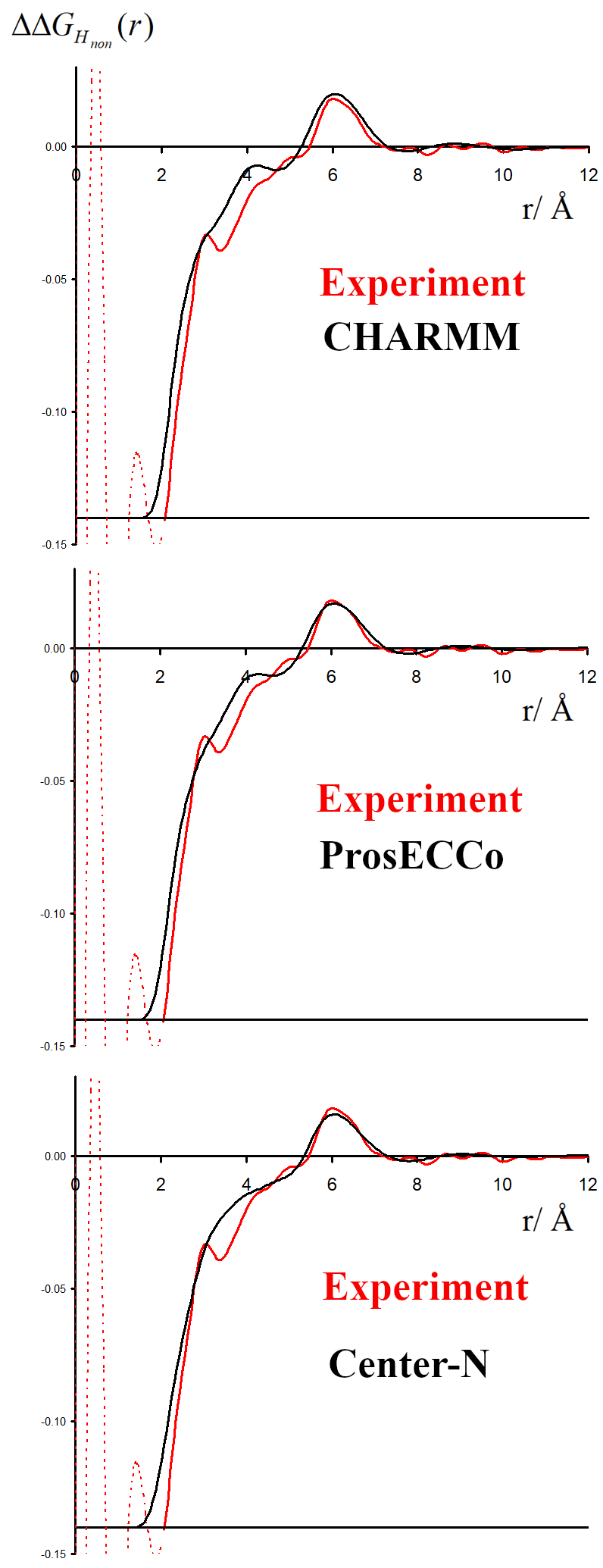}
  \caption{The function $\Delta\Delta G_\mathrm{H_{non}}(r)$ is shown in black, with the same function calculated from FFMD simulations in red for the \texttt{CHARMM} (upper), \texttt{prosECCo} (middle) and \texttt{Center-N} force fields (lower).}
  \label{fgr:2ndMDvsEXP}
\end{figure}

The double difference signal, $\Delta\Delta G_\mathrm{H_{non}}(r)$, obtained from neutron scattering experiments after Fourier transform is composed of a single radial distribution function of $g_\mathrm{H_{TMA}H_W}(r)$, which can be directly compared to the same radial distribution function computed from FFMD simulations with different force fields (Fig.~\ref{fgr:2ndMDvsEXP}). All investigated force fields (\texttt{CHARMM}: full charge, \texttt{prosECCo}: scaled, and \texttt{center-N}: +0.75 on N) capture the location of the main peak at about 6~{\AA}, even if the \texttt{CHARMM} force field seems to provide a somewhat worse fit than the two others. Small differences are visible in the low-$r$ range, but the comparison does not allow us to decide on the best force field. The steep rise of the experimental signal is slightly shifted for all force fields, and the shoulders around 4--5~\AA\, while all slightly different, are never the same as in the experimental signal.

\begin{figure}[ht!]
\centering
  \includegraphics[width=8cm]{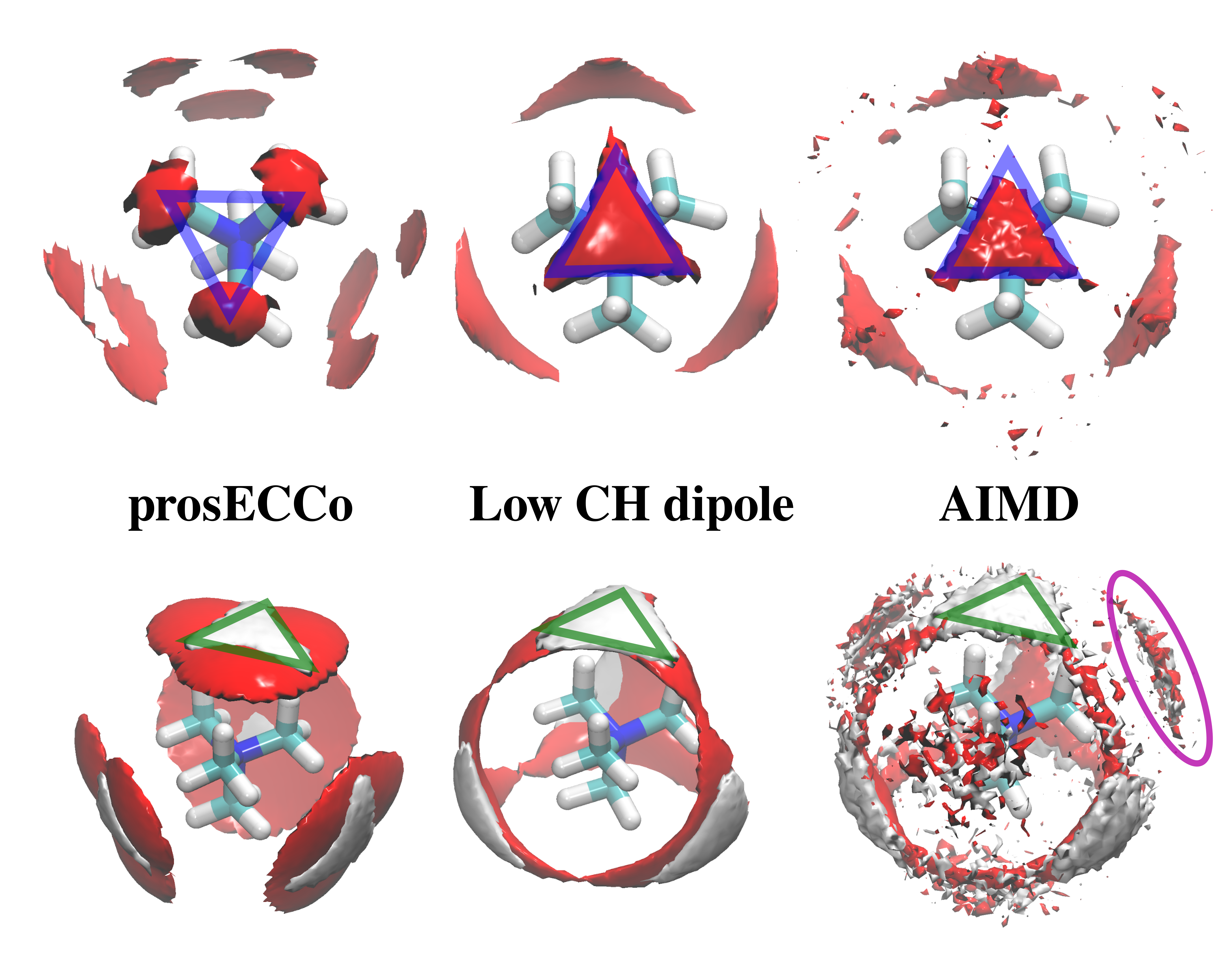}
  \caption{$\mathrm{O_{W}}$ and $\mathrm{H_{W}}$ density maps around TMA in various simulations. Upper $\mathrm{O_{W}}$ density is shown in red at 3.3x bulk density (approximated for AIMD simulations) and lower 2.8x bulk density for $\mathrm{O_{W}}$ (red) and 2.1x bulk density for $\mathrm{H_{W}}$ (white). The orientation of the $\mathrm{O_{W}}$ around TMA shows an inverse relationship between \texttt{prosECCo} and \texttt{low CH dipole} (highlighted by blue triangles). The $\mathrm{H_{W}}$ densities are very similar for \texttt{prosECCo} and \texttt{low CH dipole} FFs (highlighted by green triangles). The $\mathrm{O_{W}}$ densities for AIMD and \texttt{low CH dipole} are similar, however, none of the FFMDs replicate the $\mathrm{H_{W}}$ and $\mathrm{O_{W}}$ clouds off the corner of the TMA tetrahedron (highlighted in blue) at any density contour level.}
  \label{fig:aimd}
\end{figure}

As neutron scattering data cannot differentiate between the different 3D hydration arrangements found by FFMD, we performed DFT-based AIMD simulations and used them as a benchmark. Since the obtained FFMD density maps differ qualitatively from each other, we aimed to employ the AIMD results to determine which force field is more accurate. The very high computing cost of such simulations (see Methods for more details) necessarily limits both the size of the simulated system (a single TMA ion in a small box of 64 water molecules) and the length of the simulations (500~ps). Due to the limited box size, the calculation of radial distribution functions is thus limited to the small $r$ range (Fig.~2 in ESI). We also obtained 3D density maps for water oxygen atoms around TMA. Despite limited statistics, these plots tend to show that the highest density of water oxygen atoms is located at the center of the faces of the TMA tetrahedron, with bridges across the sides. The hydration density maps from AIMD (see Fig.~\ref{fig:aimd}) display both similarities and differences with respect to the FFMD density maps. A feature common to all density maps is that the oxygen clouds found over the faces and edges are closer to TMA than those of the $\mathrm{H_{W}}$ clouds. There is, however, a difference in how these clouds are arranged for the \texttt{prosECCo} and \texttt{low CH dipole} force fields. In each case, the $\mathrm{H_{W}}$ clouds are similar in shape but vary such that the \texttt{low CH dipole} force field has a greater tendency to spill over the edges of the TMA tetrahedron. For the \texttt{low CH dipole} force field the O clouds match the orientation of the $\mathrm{H_{W}}$ cloud, while for the \texttt{prosECCo} force field, an opposite pattern is observed with the triangles of the $\mathrm{H_{W}}$ and O cloud being anticorrelated (Figure \ref{fgr:MapsGeometry} and \ref{fig:aimd}). Hydration of the AIMD TMA is more similar to that of \texttt{low CH dipole} than that of \texttt{prosECCo}, where \texttt{prosECCo} and \texttt{CHARMM} models have similar hydrations. Interestingly, at lower atomic densities, AIMD density maps show overlapping $\mathrm{H_{W}}$ and $\mathrm{O_{w}}$ density clouds at the corners of TMA, which is not replicated in any force field-based FFMD simulation. Taken all together, these results suggest that the orientation of water molecules in the TMA hydration shell is better captured by \texttt{low CH dipole} than by the other force field variants tested here.

\begin{figure*}[ht]
\centering
  \includegraphics[width=18cm]{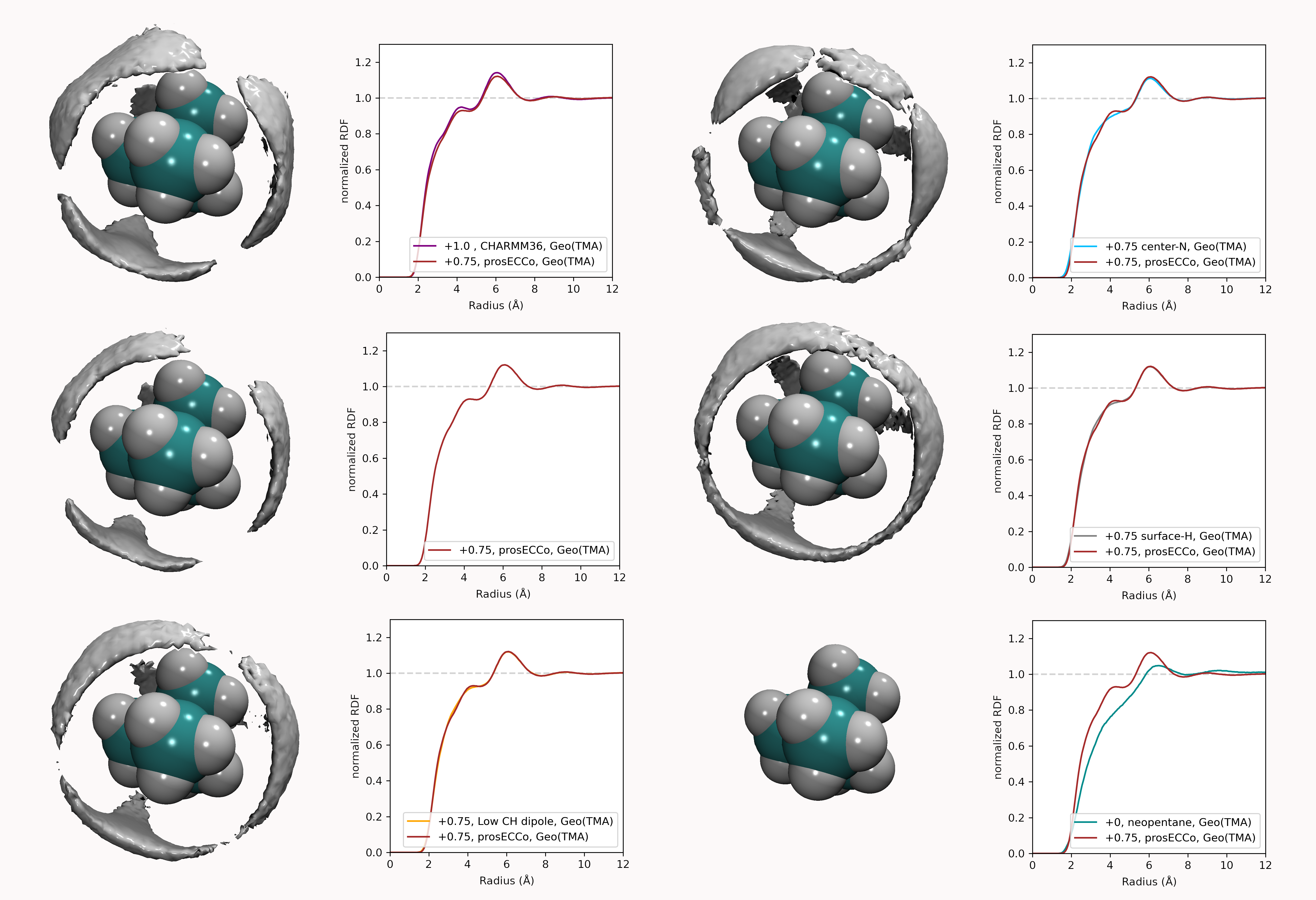}
  \caption{Density maps of water hydrogen atom around TMA, at twice bulk density. Shown to the right of each density map is the corresponding $\mathrm{H_{TMA}}$-$\mathrm{H_{W}}$ radial distribution function, compared to the same function for the \texttt{prosECCo} TMA force field to help comparison.}
  \label{fgr:MapsMDvsEXP}
\end{figure*}

To further investigate the sensitivity of the neutron scattering signal to different aspects of TMA hydration, we compared $g_\mathrm{H_{TMA}H_W}(r)$ (which is directly related to the experimentally measured quantity) among all the different variants of the TMA force fields. While we previously showed (Fig.~\ref{fgr:MapsCharge}) that different charge distributions lead to strikingly different hydration patterns around TMA and different ion-pairing behavior, resulting in different patterns visible at the $\mathrm{H_{W}}$ density map around TMA (Fig.~\ref{fgr:MapsMDvsEXP}), these differences do not significantly modify the $g_\mathrm{H_{TMA}H_W}(r)$ (see Fig.~\ref{fgr:MapsMDvsEXP}, where the $g_\mathrm{H_{TMA}H_W}(r)$ computed with different force fields are compared). Neutron scattering experiments examining the $g_\mathrm{H_{TMA}H_W}(r)$ correlation are thus unable to distinguish the different hydration patterns and ion pairing propensities that we have shown to exist for these different force fields. \textit{It is rather unexpected that these different three-dimensional water orientations around the different TMA force fields give such similar RDFs for $g_\mathrm{H_{TMA}H_W}(r)$.} The origin of the similarity of the RDFs despite the significant differences in the density maps is as follows. The hydrogen density clouds have symmetric ordering around the center of the TMA molecule. However, the substituted nuclei are not at this symmetry center, and there is a correlation between each of the clouds of $\mathrm{H_{W}}$ density and each of the 12 substituted hydrogens, which makes the function $g_\mathrm{H_{TMA}H_W}(r)$ rather broad. The $g_\mathrm{H_{TMA}H_W}(r)$ component of the structural data, despite being a large fraction of the total scattering, is thus not very informative of the relevant ion hydration structure. Even if we had examined the correlations between the central nitrogen of the TMA ($\mathrm{N_{TMA}}$) and $\mathrm{H_{W}}$, the corresponding RDFs would still be all remarkably similar to each other for all of these force fields (see Fig.~\ref{fgr:2ndMDvsEXP}), and would have produced a far smaller NDIS signal. Only, if the correlation between the central nitrogen and the oxygen in water could be measured, this would enable differentiating between these force fields. Unfortunately, isolating this component of the total scattering signal with isotopic substitution is not possible, as no suitable oxygen isotopes exist for such an experiment.

\section{Conclusions}

Neutron scattering experiments with double isotopic substitution on both $\mathrm{H_{TMA}}$ and $\mathrm{H_{water}}$ were performed allowing us to single out the correlation between $\mathrm{H_{TMA}}$ and $\mathrm{H_{water}}$. Thanks to the very high contrast of the H/D substitution and the large number of H atoms in TMA, the signal is well above the noise level. It should thus allow for a detailed characterization of the hydration structure. However, we show here that molecular interpretation of the experimental signal proves to be very challenging as all experimentally measurable RDFs ($g_\mathrm{H_{TMA}H_W}(r)$ and $g_\mathrm{NH}(r)$) are not very sensitive to different hydration patterns caused by changes in the employed force fields. We thus can neither fully infer the most probable hydration structure from such a comparison nor validate the preferential choice of the tested force field variants. 
Nevertheless, the simulation results provide important insights into TMA hydration and its sensitivity to force field parameters. Shifting from \texttt{CHARMM} (+1.0) to \texttt{prosECCo} (+0.75) charges had a relatively small effect on the ion hydration. Changing the polarity of the C--H bond from $-$0.35 C and +0.23 H (\texttt{prosECCo}) to $-$0.1 and +0.1 (\texttt{low CH dipole}) largely inverts the hydration structure of oxygens around the TMA. AIMD simulations show a hydration structure very similar to \texttt{low CH dipole} at the faces and edges of the TMA tetrahedron. The AIMD hydration structure seen at the corners of the TMA tetrahedron is, however, not replicated by any FFMD. Thus, we suggest that the C--H bond polarity in standard \texttt{CHARMM} and its variants is too high to capture the TMA hydration properly.
Removal of the tetrahedral structure by employing a \texttt{center-bead} force field for TMA and converting it to a single large bead has a relatively minor effect on the radial hydration structure of the ion. Still, it strongly changes the form of the counterion interaction. Namely, our findings imply that longer-range ion--ion ordered structures may not be accurately replicated as the structure of TMA is simplified to the level of a single bead within a coarse-grained force field.
All these results suggest that caution should be taken when simulating moieties with TMA groups where hydration may play a relevant role, such as for common phospholipids and methylated lysines extensively present in biological systems.


\section*{Conflicts of interest}
There are no conflicts to declare.

\section*{Acknowledgements}
H.M.-S. acknowledges the support of the Czech Science Foundation (project 19-19561S).
O.T. acknowledges the Faculty of Mathematics and Physics of the Charles University (Prague, Czech Republic) where he is enrolled as a PhD student, and the International Max Planck Research School for ``Many-Particle Systems in Structured Environments'' (Dresden, Germany) for support. M. V. acknowledges support by the Ministry of Education, Youth and Sports of the Czech Republic through the e-INFRA CZ (ID:90254), Project OPEN-28-18. EDD acknowledges support from the “Initiative d’Excellence” program from the French State (Grant “DYNAMO”, ANR11-LABX-0011). P.J. acknowledges support from the European Research Council via an ERC Advanced Grant no. 101095957.



\balance


\bibliography{library} 
\bibliographystyle{rsc} 

\pagebreak
\begin{center}
\noindent\LARGE{\textbf{ELECTRONIC SUPPLEMENTARY INFORMATION (ESI):}} \\ 
\noindent\LARGE{\textbf{Hydration of biologically relevant tetramethylammonium cation by neutron scattering and molecular dynamics}} \\
\end{center}
\setcounter{equation}{0}
\setcounter{figure}{0}
\setcounter{table}{0}
\makeatletter
\renewcommand{\theequation}{S\arabic{equation}}
\renewcommand{\thefigure}{S\arabic{figure}}
\renewcommand{\bibnumfmt}[1]{[S#1]}
\renewcommand{\citenumfont}[1]{S#1}

\section*{Additional details on neutron scattering}

Measuring a total neutron scattering pattern for a solution yields an insight into the correlations between every atom in the system with every other atom in the system. While this is comprehensive, for most multi-atom systems it is too complex to interpret. Neutron scattering with isotopic substitution allows components of this to be isolated and examined separately. In the main body of the paper, we isolate the H\textsubscript{TMA}-H\textsubscript{W} component.  However, other partial structure factors can be isolated from this experimental data. The most obvious to look at is the correlations between the H\textsubscript{TMA} and all nuclei in the system that are not H\textsubscript{TMA}. This is achieved by subtracting the H\textsubscript{TMA}-H\textsubscript{W} component from the first-order difference in the main body of the paper and is summarized in Eq.~\ref{eqn:Q1stH2OY} in ESI. This is shown in real and reciprocal space in Fig.~\ref{eqn:Q1stH2OY} in ESI. While a principal component of this function is the correlation between H\textsubscript{TMA} and O\textsubscript{W} on the solvating water, in practice it is very difficult to gain any useful insight from this function.  This is because the intramolecular correlations between H\textsubscript{TMA} and the other atoms on the TMA give very sharp peaks due to the molecular bonds. These very large peaks lie directly on top of the most interesting part of the solvation of the TMA ion in the 3--4~\AA~range.

\begin{equation}
  \label{eqn:Q1stH2OY}
  \begin{split}
  \Delta S\mathrm{^Y_{H_{non}}}(Q) & = 39.3 \cdot S\mathrm{_{H_{TMA}O}}(Q) + 6.5 \cdot S\mathrm{_{H_{TMA}C}}(Q) \\
                                & + 2.3 \cdot S\mathrm{_{H_{TMA}N}}(Q) + 2.3 \cdot S\mathrm{_{H_{TMA}Cl}}(Q) \\
                                & + 4.3 \cdot S\mathrm{_{H_{TMA}H_{TMA}}}(Q) - 54.6
  \end{split}
\end{equation}

\noindent where $\mathrm{Y}$ refers to every atom in the system but exchangeable hydrogen atoms. This equation can be Fourier transform to the real space as:

\begin{equation}
  \label{eqn:G1stH2OY}
  \begin{split}
  \Delta G\mathrm{^Y_{H_{non}}}(r) & = 39.3 \cdot g\mathrm{_{H_{TMA}O}}(r) + 6.5 \cdot g\mathrm{_{H_{TMA}C}}(r) \\
                                & + 2.3 \cdot g\mathrm{_{H_{TMA}N}}(r) + 2.3 \cdot g\mathrm{_{H_{TMA}Cl}}(r) \\
                                & + 4.3 \cdot g\mathrm{_{H_{TMA}H_{TMA}}}(r) - 54.6.
  \end{split}
\end{equation}

\begin{figure}[H]
\centering
  \includegraphics[width=8cm]{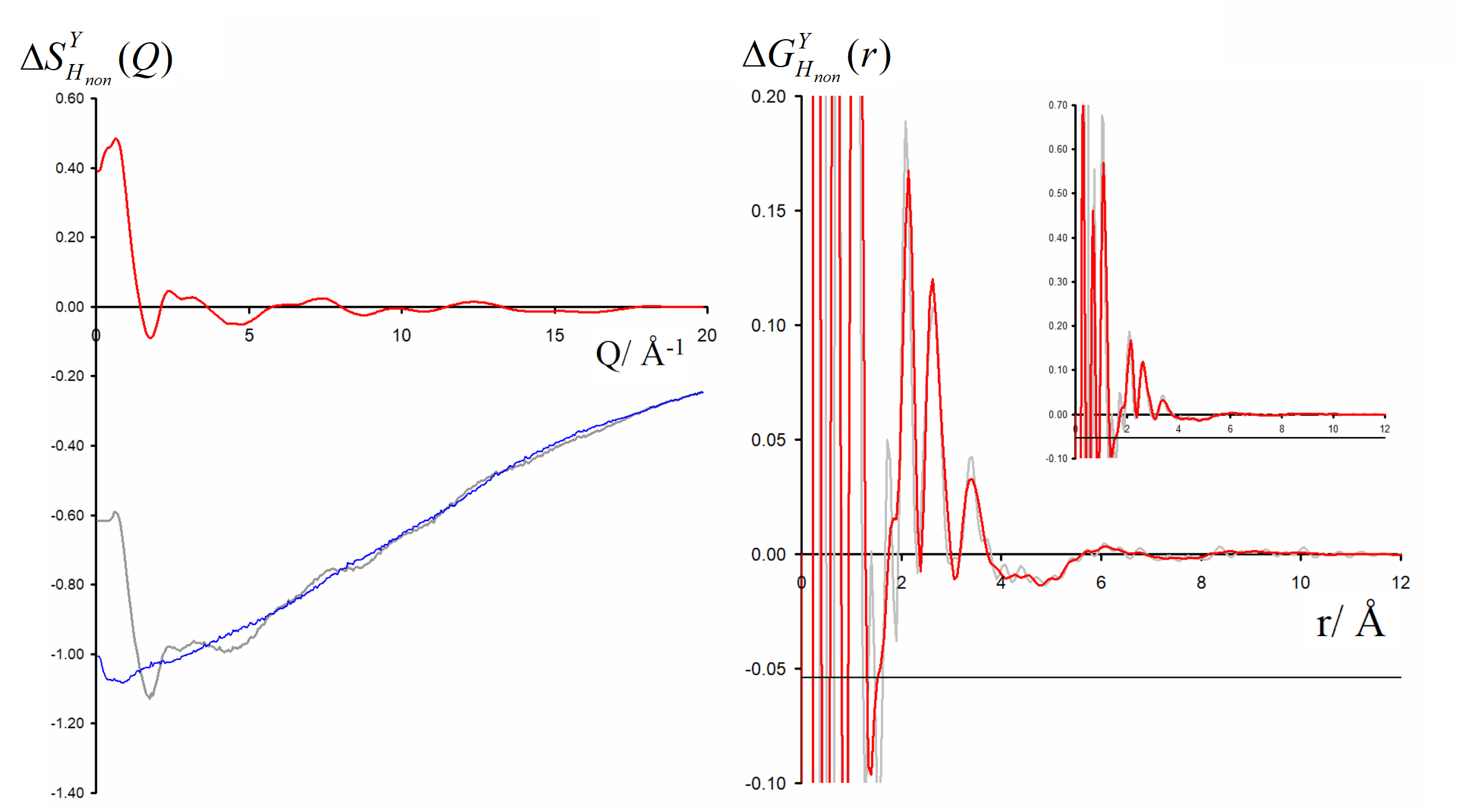}
  \caption{Left in grey is the raw function $\Delta S\mathrm{^Y_{H_{non}}}(Q)$. In the right, in grey is the direct Fourier transform of this function, $\Delta G\mathrm{^Y_{H_{non}}}(r)$. The function $\Delta S\mathrm{^Y_{H_{non}}}(Q)$ after background subtraction and smoothing is shown in red. In the right, in red is its direct Fourier transform $\Delta G\mathrm{^Y_{H_{non}}}(r)$. As the H\textsubscript{TMA}-C intramolecular peak of the TMA at 1.1~\AA~is not very interesting in $\Delta G\mathrm{^Y_{H_{non}}}(r)$ and dwarfs everything else in this function, this is only shown in the inset.}
  \label{fgr:1stY}
\end{figure}

\section*{Extra data AIMD}

Contrary to all tested force fields, the AIMD simulation captures the shoulder in the experimental signal at low $r$. However, the box is too small to assess the signal quality at the position of the main peak and beyond. Still, the small size of the box may affect the RDF. The reasonable agreement with the experimental signal is obtained despite the AIMD simulation not containing any chloride counterion. 

\begin{figure}[H]
\centering
  \includegraphics[width=6.0cm]{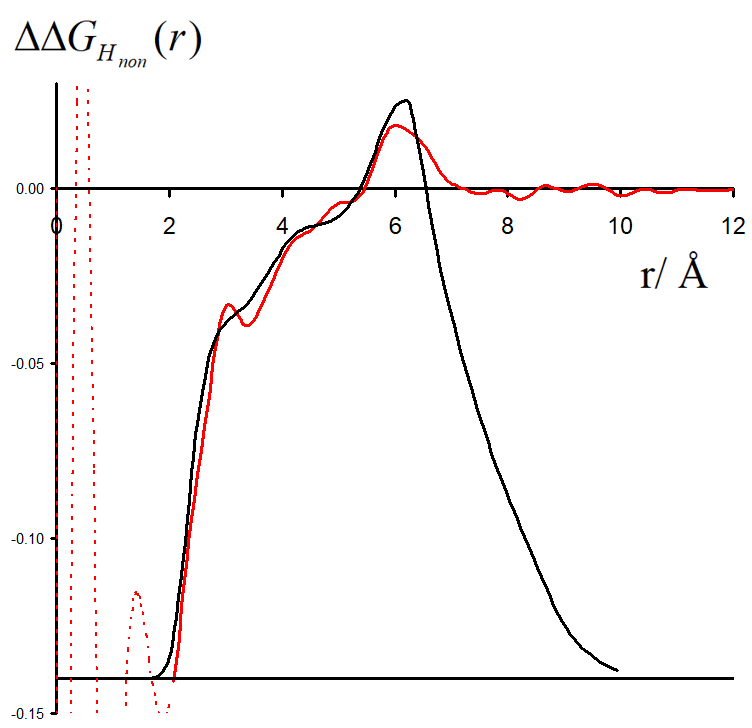}
  \caption{Experimental Fourier-transformed signal $\Delta\Delta G_\mathrm{H_{non}}(r)$ (red), together with the same $\Delta\Delta G_\mathrm{H_{non}}(r)$ obtained from AIMD simulations (black).}
  \label{fgr:QM}
\end{figure}

\end{document}